\documentclass[
 reprint,
 amsmath,amssymb,
 aps,
prx
]{revtex4-2} 
\renewcommand{\selectlanguage}[1]{}

\usepackage{comment}
\usepackage{graphicx}
\usepackage{dcolumn}
\usepackage{bm}

\usepackage{placeins}

\usepackage[hidelinks]{hyperref}
\usepackage{xcolor}

\usepackage[normalem]{ulem}

\usepackage{siunitx}

\usepackage{booktabs}

\newcommand{\txtmath}[1]{$\mathrm{#1}$}

\newcommand{\hen}{He$\mathrm{_N}$}
\newcommand{\MgExcited}{$\mathrm{3^1P_1}$~}
\newcommand{\MgGround}{$\mathrm{3^1S_0}$~}

\newcommand{\TUG}{Institute of Experimental Physics, Graz University of Technology, Petersgasse 16, 8010 Graz, Austria.}

\newcommand{\foamdecaytime}{$\tau_2=(4.0 \pm 0.9)$\,ps} 
\newcommand{\foamrisetime}{$\tau_2^\textrm{rise}=(450 \pm 180)$\,fs}
\newcommand{\compactdecaytime}{$\tau_1=(380 \pm 70)$\,fs} 
\newcommand{\atomdecaytime}{$\tau_3=(1.1 \pm 0.1)$\,ps} 

\newcommand{\splitenergy}{$0.42$}

\begin{document}

\preprint{APS/123-QED}

\title{ Real-time tracking the energy flow in cluster formation}

\author{Michael Stadlhofer}
 \altaffiliation[\TUG]{}
  \author{Bernhard Thaler}
   \altaffiliation[\TUG]{}
 \author{Pascal Heim}
  \altaffiliation[\TUG]{}
\author{Josef Tiggesbäumker}
    \altaffiliation[Institute of Physics, University of Rostock, 18059 Rostock, Germany]{}
    \altaffiliation[Department of Life, Light and Matter, University of Rostock, 18059 Rostock, Germany]{}
\author{Markus Koch}
 \altaffiliation[\TUG]{}
 \email{markus.koch@tugraz.at}

\date{\today}

\begin{abstract}

While photodissociation of molecular systems has been extensively studied over decades, the photoinduced formation of chemical bonds remains largely unexplored. Especially for larger aggregates, the electronic and nuclear dynamics involved in the cluster formation process remain elusive. This limitation is rooted in difficulties  to prepare reactants at well-defined initial conditions. 
Here, we show that this hurdle can be overcome by exploiting the exceptional solvation properties of superfluid helium  nanodroplets (\hen).
We load the droplets with magnesium (Mg) atoms and investigate the dynamical response of the formed Mg$_\textrm{n}$ aggregates to photoexcitation with femtosecond time-resolved photoelectron spectroscopy. 
Beside the response expected for conventional  Mg$_\textrm{n}$ clusters, consisting of a prompt signal rise and a decay characteristic for van der Waals bonds, the transient spectra  also show a delayed photoelectron band peaking at 1.2\,ps.
This delayed signal rise is characteristic for nuclear dynamics and represents the photoinduced transition of Mg$_\textrm{n}$ aggregates from a metastable, foam-like configuration, where Mg atoms are stabilized with a previously predicted interatomic spacing of 9.5\,\AA, to a compact cluster. 
With global fitting analysis and photoion-photoelectron coincidence detection, the concerted electronic and nuclear dynamics can be tracked on a fs timescale. We find that cluster formation,  proceeding with a characteristic time constant of ($450\pm180$)\,fs, is accompanied by the population of highly-excited atomic states, exceeding the photoexcited state by up to 3\,eV. 
We propose an energy pooling reaction in collisions of two or more excited Mg atoms during cluster formation as the mechanism leading to population of these high-lying Mg states. 
  Additionally, conversion to kinetic energy through relaxation of the highly-excited states leads to fragmentation and enables ionic cluster fragments to overcome the He droplet solvation energy. 
These results underline the potential of \hen ~for time-resolved studies of bond formation and to uncover involved processes, such as photon energy upconversion.

\end{abstract}

\maketitle

\section{Introduction}
Chemical reactions essentially consist of breaking and forming of molecular bonds.
The mechanistic understanding of photoinduced bond breaking has been particularly shaped through femtosecond pump--probe spectroscopy.~\cite{Weinacht2019,Stolow2004,Hertel2006} Real-time tracking of the electronic and nuclear structure has provided insight into  various  processes accompanying photodissociation, such as 
curve crossings~\cite{Kobayashi2019,Arasaki2003}, 
predissociation~\cite{Mokhtari1990}
conical intersections~\cite{Nunn2010}, or
electronic relaxation~\cite{Koch2017}.

Photoassociation is used to form molecules in ultracold atomic clouds  \cite{Fioretti1998, Jones2006}, where the binding energy is dissipated into the light field \cite{Vitanov2001}. 
In the time--domain, however, photoassociation has largely escaped observation so far, in particular for larger systems, due to difficulties in preparing the reactants in a well-defined initial geometry.
This limitation could be overcome only in a few situations: 
In gas phase, where the broad distribution of impact parameters completely blurs the time resolution, bond formation could only be observed for selected dimer molecules through Franck--Condon filtering based on the resonance condition for laser excitation~\cite{Marvet1995,Rybak2011a,Rybak2011}. In a few cases, cold bimolecular van der Waals complexes~\cite{Scherer1987,Gruebele1991,Potter1992a,Stert2001a}, and anionic clusters~\cite{Wester2003}, provide favorable initial conditions to study bond formation. 
In solution, a network of species separated by well-defined distances are readily achieved, however, inhomogeneous broadening prevents observation of individual states and structural dynamics can only be inferred by x-ray scattering~\cite{Kim2015,Lee2013}.
Despite these versatile approaches, the formation of clusters larger than dimer molecules and bimolecular aggregates has not been experimentally realized so far.
\\
\begin{figure*}
    \centering
    \includegraphics[width=\linewidth]{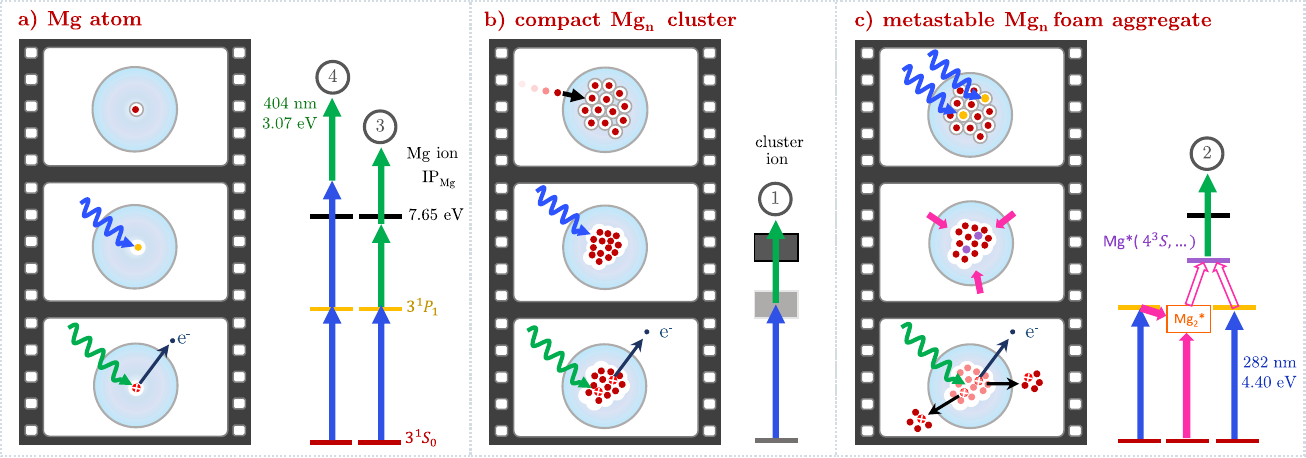}
    \caption{Sketch of the photoinduced dynamics of single Mg atoms, compact Mg$_\textrm{n}$ cluster and metastable Mg$_\textrm{n}$ foam inside \hen. The corresponding energy level diagrams depict pump (blue)--probe (green) photoionization and different ionization pathways are labelled with (1) to (4), correspondingly to the photoelectron bands in Fig.~\ref{fig:timescan_all_electrons}.
\textbf{(a)} Three-photon ionization channels of Mg atoms. 
\textbf{(b)} Formation of compact Mg$_\textrm{n}$ cluster through collision of an energetic Mg atom with the foam-like aggregate during the pickup process~\cite{Kazak2022} and subsequent pump--probe photoionization.
\textbf{(c)} Photoexcitation of Mg atoms within the Mg$_\textrm{n}$ foam triggers the transition to a compact cluster.
Energy pooling (magenta arrows) leads to population of highly excited Mg levels [band (2)] and fragmentation.
}
    \label{fig:schematic}
\end{figure*}
The endeavor to combine the advantages of the previous approaches---homogeneous distances in solution,  low environmental perturbation for well-defined resonances, and low temperatures in gas phase---leads us to the application of helium nanodroplets (\hen) as a cryogenic solvent.
Helium nanodroplets have routinely been used for the synthesis and investigation of atomic and molecular aggregates, since they provide a high degree of control in the aggregation process, efficient cooling to 0.37\,K, and enable measurements with low matrix effects compared to other noble gas environments~\cite{Toennies2004,Callegari2011,Ernst2021,Slenczka2022}. 
Here, we show that the  unique solvation properties of superfluid He can be used to prepare well-defined initial conditions for the time-resolved observation of bond formation of multiple reactants.
This demonstration builds on recent observations  that atoms solvated in \hen~can  arrange in metastable configurations at nanometer distance, enabled by the ultracold He solvent~\cite{Albrechtsen2023, Kazak2019, Lackner2018,  Kautsch2015, Goede2013, Poms2012,Przystawik2008}.
For two Mg atoms inside a \hen, density functional theory simulations predict such a metastable configuration at 9.5\,\AA\ interatomic distance, instead of the formation of a Mg$_2$ molecule~\cite{Hernando2008}. 
This dilute configuration is enabled through the accumulation of helium density between two Mg atoms, and referred to as ``foam'' ~\cite{Hernando2008} or ``quantum gel''~\cite{Eloranta2008,Eloranta2011}.
We note that this prediction has been subject to discussion, since path integral Monte Carlo simulations of two and three Mg atoms inside \hen~find no evidence for stabilization in a metastable configuration, but only equilibration to the strongly bound dimer and trimer~\cite{Krotscheck2016}.
\\
Here, we present an  investigation of the dynamical response of Mg$_\textrm{n}$ aggregates inside \hen ~to photoexcitation with femtosecond photoelectron and -ion spectroscopy.~\cite{Stolow2004,Hertel2006}. This approach has recently proven successful inside \hen~for the observation of electronic~\cite{Thaler2018,Bruder2022} and nuclear dynamics~\cite{Thaler2020a, Nielsen2022}.
    We obtain time-resolved photoelectron spectra (TRPES) containing two different dynamic signatures, indicating that Mg$_\textrm{n}$ aggregates exist in two different configurations inside \hen. An immediate signal rise followed by a fast \compactdecaytime ~decay is characteristic for van der Waals cluster, where primary processes are electronic dynamics (Fig.~\ref{fig:schematic} b).
In addition to this swiftly responding compact cluster, the TRPES also contains a PE band with delayed signal rise peaking 1.2\,ps after photoexcitation and a much slower \foamdecaytime ~decay. We interpret this slower response to photoexcitation as a signature for nuclear dynamics involved in the transition leading from the predicted foam-like Mg$_\textrm{n}$ configuration to a dense cluster (Fig. \ref{fig:schematic}c). 
This transient signal reveals insight into the energy flow and nuclear dynamics during cluster formation.

\section{Experiment}\

Following a previous approach~\cite{Thaler2018},  helium nanodroplets with a mean radius of  5.3\,nm (13500 He atoms per droplet) are loaded with about ten Mg atoms (see Supplemental \ref{subsec:doping} for further details). 
  Using an amplified Ti:sapphire laser   (800\,nm center wave length, 25\,fs pulse duration), short pulses are generated and split into a pump and probe arm. The cross-correlation signal of the pump and probe pulse has a duration of $(45\pm3)$\,fs. 
The pump pulse is tuned to 282\,nm (4.40\,eV photon energy) by an optical parametric amplifier, in order to trigger the collapse of the foam-like \txtmath{Mg_n} aggregate through Mg \MgExcited$\leftarrow$\MgGround excitation, which appears slightly blue-shifted in the aggregate~\cite{Przystawik2008} relative to the bare atom transition~\cite{NIST_ASD} (see Fig.~\ref{fig:schematic}a). 
The probe pulse at 404\,nm (3.07\,eV), obtained through frequency doubling,  ionizes the system and photoelectron spectra are recorded with a magnetic bottle time-of-flight spectrometer.
The time-resolved variation of these photoelectron spectra, recorded through variation of the pump--probe time delay, provides insight into the evolution of excited state populations.
Applying a high voltage pulse to the repeller electrode about 100\,ns  after the laser pulses accelerates the remaining ions towards the detector and thus allows for a simultaneous detection of electrons and ions in each laser shot~\cite{Koch2017b}. This procedure  enables a statistical analysis of covariances between electron energy and ion species.~\cite{Frasinski1989,Mikosch2013b,Mikosch2013c}.
\section{Results}
\begin{figure}[!ht]
\includegraphics[width=\linewidth]{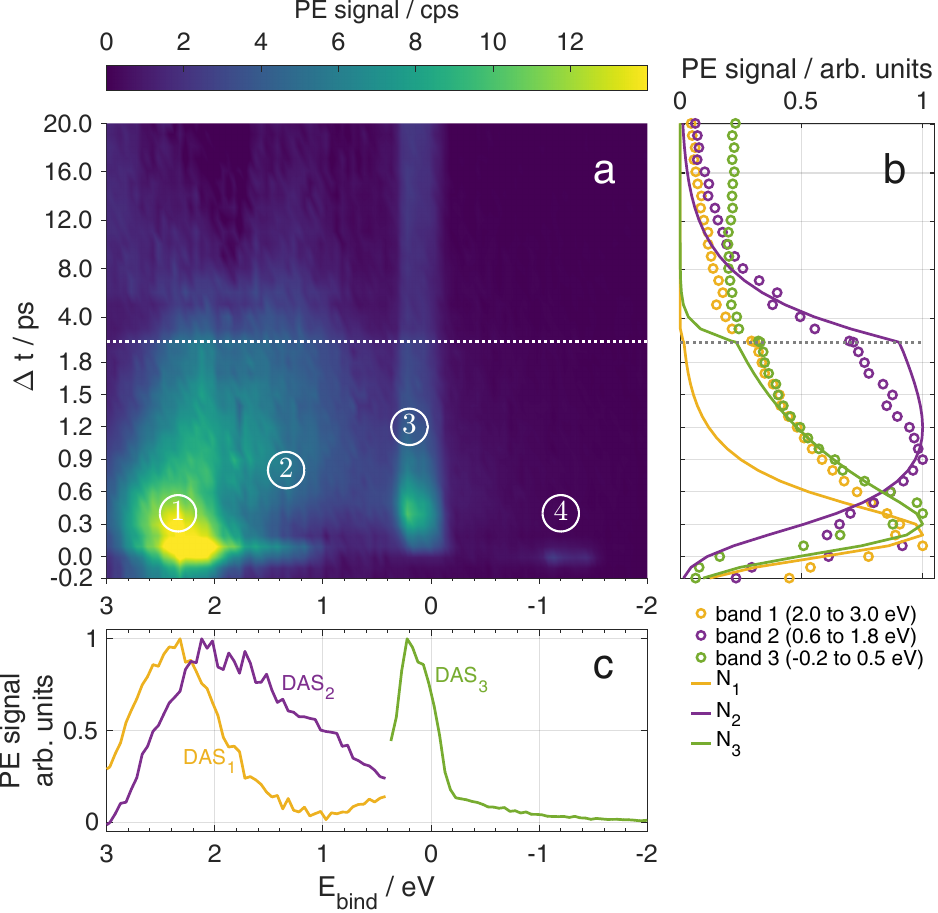}
\caption{ 
Time-resolved photoelectron spectrum of Mg$_\textrm{n}$ aggregates  embedded in  \hen~and global fitting analysis. Note that below 2\,ps, the spectra feature a higher temporal resolution.
\textbf{(a)} Pseudocolor plot showing the transient electron yield as function of binding energy. Four distinct bands are marked (1) to (4).
\textbf{(b)} Comparison of energy-integrated PE transients (see legend) to decay functions N$_{1,2,3}$ from global fitting. 
\textbf{(c)} Decay associated spectra DAS$_{1,2,3}$ obtained from global fitting. 
}
\label{fig:timescan_all_electrons} 

\end{figure}

The time-resolved photoelectron spectrum is shown in Fig.~\ref{fig:timescan_all_electrons},  depicted as a function of the electron binding energy. 
We first allocate the observed photroelectron bands to pump--probe ionization pathways based on energetic considerations. The temporal development of these bands, obtained from a global fitting analysis, reveals  population transfers dynamics triggered by photoexcitation.
Examination of the time-resolved ion yields and, in particular, the correlation of ion fragments to the photoelelctron bands gives information about the accompanying nuclear dynamics.\\
\\
\textbf{Assignment of photoelectron bands.}
Inspection of Fig.~\ref{fig:timescan_all_electrons}a reveals four distinct  bands.
Band (1) between 2 and 3\,eV rises instantaneously and shows a fast decay  (Fig.~\ref{fig:timescan_all_electrons}b, yellow circles). 
Band (2) between 1 and 2\,eV shows initially a cross correlation feature, followed by a moderate signal rise peaking at 1.2\,ps and a slower decay than band (1) (Fig.~\ref{fig:timescan_all_electrons}b, purple circles). 
The assignment of bands (1) and (2) in Fig.~\ref{fig:timescan_all_electrons} is difficult due to their overlap in time and energy. 
Both bands are related to states with binding energies lower than that of the photoexcited bare-atom \MgExcited state (3.30\,eV binding energy), given the lower probe photon energy of only 3.07\,eV. 
Bands (3) and (4) can be assigned to two-photon ionization of the excited \MgExcited Mg state (see Fig.~\ref{fig:schematic}a). Band (3) extends from 0.17~eV to -0.5\,eV with a fast rise and slower decay, followed by a second slow  rise after 10\,ps (see Fig.~\ref{fig:timescan_all_electrons}b). 
This band represents the transient population of \MgExcited Mg state, ionized by two probe photons. 
The brief appearance of band (4) around $t_0$ is characteristic for a cross-correlation peak: With a binding energy slightly below --1\,eV we assign this band to \MgExcited ionization with one pump and one probe photon. \\

\begin{table}[h]
    \centering
    \caption{Relevant parameters of the three decay functions as obtained from the global fitting procedure (see Appendix~\ref{subsec:Supplement:Global Fitting Method} for formulas of the decay functions).  Uncertainties represent a confidence level of 95\%.  In addition to the parameter and its value, the rightmost column indicates the DAS corresponding to the decay function.
     }
    \centering
    \setlength{\tabcolsep}{8pt} 
    \begin{tabular*}{\linewidth}{@{\extracolsep{\fill}}ccl@{}}
    \toprule
    \textbf{parameter}& \textbf{value} / fs & \textbf{feature} \\ 
    \hline 
    time zero, $t_0$& $45\pm4$ & all \\ 
    temporal instrument response, $\sigma$ & $170\pm15$ & all \\ 
    $N_1$ decay time, $\tau_1$ & $380\pm70$ & DAS$_1$ \\
    $N_2$ rise time, $\tau_{2}^{rise}$ & $450\pm180$ & DAS$_2$ \\ 
    $N_2$ decay time, $\tau_{2}$ & $4000\pm900$ & DAS$_2$ \\ 
    $N_3$ decay time, $\tau_3$ & $1090\pm90$ & DAS$_3$ \\ 
    \bottomrule
    \end{tabular*}
    \label{tab:L2_parameters}
\end{table}

\textbf{Global fitting analysis to retrieve population transfer dynamics.}
The spectral and temporal overlap of bands (1) and (2) in the time-resolved photoelectron spectrum (Fig.~\ref{fig:timescan_all_electrons}) poses a challenge for a quantitative analysis.
In order to decode the different contributions,  the spectrum is analyzed by applying a global fitting procedure~\cite{Stokkum2004,Wu2011a}, i.e., 
species with different transient behavior contributing to the spectrum are identified by 
extracting the respective decay associated spectra, DAS($E$), and the associated transient decay functions, $N(t)$. 
Since the photoelectron bands in Fig.~\ref{fig:timescan_all_electrons} differ in their signal rise behavior, we use two different types of decay functions: An instantaneous signal rise is modeled by a directly excited state followed by population decay (a Gaussian function convoluted with an exponential decay, as described in detail in Appendix~\ref{subsec:Supplement:Global Fitting Method}). 
A delayed signal rise is 
modeled by assuming sequential population transfer from an initially excited state (which is not necessarily detected) into the state yielding the photoelectron signal of the observed band. This function accounts for an exponential signal rise followed by an exponential decay, with two different characteristic time constants.
\\
Taking into account photoelectron-band assignments already made, one can simplify the fit process:  Since band (3) originates from a different ionization process than bands (1) and  (2) and since there is no overlap of band (3) with bands (1) and (2), one can split the time-resolved PE data into two energy domains at \splitenergy~eV binding energy. 
The low binding energy region, containing the \MgExcited atom band (3), is modeled by \txtmath{DAS_3} (Fig.~\ref{fig:timescan_all_electrons}c, green line). The \MgExcited population appears instantly (Fig.~\ref{fig:timescan_all_electrons}b, green line), suggesting direct excitation by the pump pulse, and then decays with a time constant of \atomdecaytime . 

Relevant parameters of the decay functions are listed in Tab.~\ref{tab:L2_parameters} for better comparability.
From this transient signal, one can deduce the temporal pump-probe overlap (time zero, $t_0$) and the temporal instrument response function duration $\sigma$.  
Inspection of the low binding energy region in Fig.~\ref{fig:timescan_all_electrons} shows that the signal rises towards long delay times, which is not represented for by the fit function.
We account for this deviation by introducing an additional background, as described in Appendix~\ref{subsec:Supplement:Global Fitting Method}.
Furthermore, the pump--probe cross correlation signal of  band (4) is determined to be  $(45\pm3)$\,fs in a separate measurement of the  the total electron yield around time zero.
This cross correlation signal, together with $t_0$ and $\sigma$, is kept constant in the remaining global fitting process.
\\
In the high binding-energy domain two distinct populations, represented by different decay associated spectra and different decay functions, can be expected. 
In agreement with this assumption, \txtmath{DAS_1} with a peak at 2.5\,eV  (Fig.~\ref{fig:timescan_all_electrons}c, yellow line) and the broader $\mathrm{DAS}_2$ peaking at 2\,eV (purple line) can be identified, together with two corresponding decay functions. The transient population $N_1$ rises quickly to a maximum at $\sim250$\,fs, followed by a rapid decay with a characteristic time of \compactdecaytime ~(Fig.~\ref{fig:timescan_all_electrons}b, yellow line).  
$N_2$ features a delayed onset with respect to $N_1$ with a rise time of \foamrisetime ~leading to a maximum at 1.2\,ps and also a slower decay time constant of \foamdecaytime ~(Fig.~\ref{fig:timescan_all_electrons}b, purple line). 
\\

\textbf{Interpretation of DAS populations.}
The different transient behavior of bands (1) and (2) indicates that they represent two different species. The sudden appearance of band (1) represents direct electronic excitation, presumably of compact Mg$_\textrm{n}$ clusters. These compact cluster have a reduced electron binding energy compared to Mg atoms~\cite{Kazak2022} and can thus be ionized by the probe pulse immediately after excitation (see Fig.~\ref{fig:schematic}b).
The picosecond delay in the onset of band (2), in contrast, suggests that nuclear dynamics are involved, which are caused by the population of an undetected excited state. A possible origin would be the presence of a previously suggested foam-like Mg$_\textrm{n}$ configuration, in combination with  a transition to a compact Mg$_\textrm{n}$ cluster triggered through \MgExcited$\leftarrow$\MgGround photoexcitaiton by the pump pulse (see Fig.~\ref{fig:schematic}b).~\cite{Przystawik2008,Goede2013, Hernando2008}
Given the \MgExcited excited state binding energy of 3.30\,eV, it is remarkable that DAS$_2$ indicates transient population within the whole detection window given by the 3.07\,eV photon energy of the probe pulse. 
In this scenario, the occupation of these highly excited states is caused by the transition of Mg$_\textrm{n}$ aggregates from  a dilute to a compact configuration.
Testing the hypothetical assignment of band (1) to a compact Mg$_\textrm{n}$ cluster and band (2) to a forming Mg$_\textrm{n}$ cluster,  the dependence of the photoelectron signal on the Mg doping level (see Supplemental Fig.~\ref{fig:pump_probe_pickup_scan}) was investigated. At high doping, band (2) decreases and band (1) increases in relative strength, consistent with earlier studies of the spontaneous collapse of the dilute Mg aggregate at high Mg doping concentrations \cite{Kazak2022}.
\\
Direct photoexcitation of Mg atoms to states with such low binding energy, as an alternative explanation for the observed high binding energies,  can be excluded since two-photon excitation leads to ionization (4.4\,eV photon energy, 7.65\,eV ionization energy~\cite{NIST_ASD}). Also, combined excitation with one pump and one probe photon can be excluded since this can only occur when pump and probe pulse overlap.
\\

\begin{figure}
    \centering
    \includegraphics[width=\linewidth]{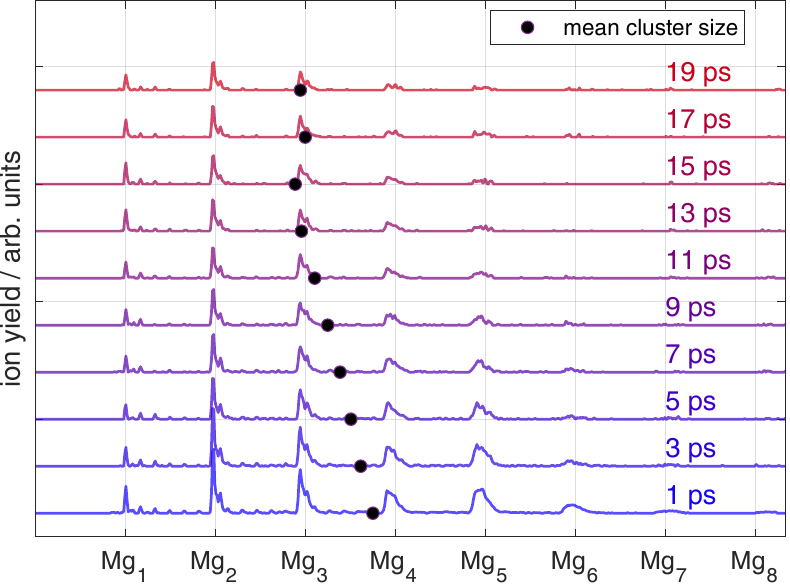}
    \caption{Ion mass spectra for different  pump--probe time delays, representing the transient fragment distribution of  Mg$_\textrm{n}^+$ cluster ions formed after photoexcitation and probe ionization. The spectra are area normalized and vertically offset according to the time-delay, indicated for each spectrum on the right side. Both isolated Mg$^+_\textrm{n}$ cluster and Mg$^+_\textrm{n}$-He$_\textrm{m}$ snowballs are present in the mass spectra. The spectra were obtained by averaging the ion mass spectra in 2~ps time intervals and the mean cluster size is indicated as a black dot. 
    }
    \label{fig:timescan_all_ions_waterfall}
\end{figure}
   
\textbf{Transient photoion signal.}
A more direct insight into the nuclear dynamics and in particular the fragmentation behavior can be gained from the ion signals.
Ions expelled from the droplet show up at optical delays larger than $\Delta t\ge1$\,ps. Figure \ref{fig:timescan_all_ions_waterfall} shows pump-probe ion spectra recorded at selected time-delays. The spectra consist of peaks corresponding to \txtmath{Mg^+_n} ($n=1-8$) cluster and \txtmath{Mg^+_n He_m} snowballs. 
\\
The pump-probe spectra show ion signals only up to \txtmath{Mg_8^+}, while in the probe-only signal, masses as high as \txtmath{Mg_{12}^+} appear. 
The average cluster size, taken as a measure and indicated by a black dot in Fig.~\ref{fig:timescan_all_ions_waterfall} for each spectrum, decreases from 3.7 to 2.8 with increasing delay.
Neglecting that the ionization probability may change with  cluster size, this down-shift reflects the general trend that an increasing amount of energy is transferred to nuclear degrees of freedom with time.
\\
\\
\textbf{Correlated electron--ion detection to identify fragmentation dynamics. }
To further test the assignment of TRPES band (1) to compact $\mathrm{Mg_n}$ cluster and band (2) to foam-like aggregates,  we examine correlations between photoelectrons and ions through covariance detection.~\cite{Frasinski1989,Mikosch2013b,Mikosch2013c} 
The assignment of ion fragments to each of the bands is of particular interest.
This detection method has been applied to gas-phase molecules, where it allows to distinguish between different photochemical reaction pathways based on the ionic products~\cite{Couch2017,Maierhofer2016,Wilkinson2014}. For photoionization
inside a helium droplet, the electrostrictive attraction prevents ion detachment from the droplet, except for situations where the ions gain sufficient kinetic energy through photodissociation~\cite{Stadlhofer2022,Braun2007}, or Coulomb explosion~\cite{Nielsen2022,DoePRL05}.  In our case, we will see that the processes behind photoelectron band (1) and (2) differ significantly in the probability to yield $\mathrm{Mg_n^+}$ ion ejection from the droplet.
\begin{figure}
\includegraphics[width=\linewidth]{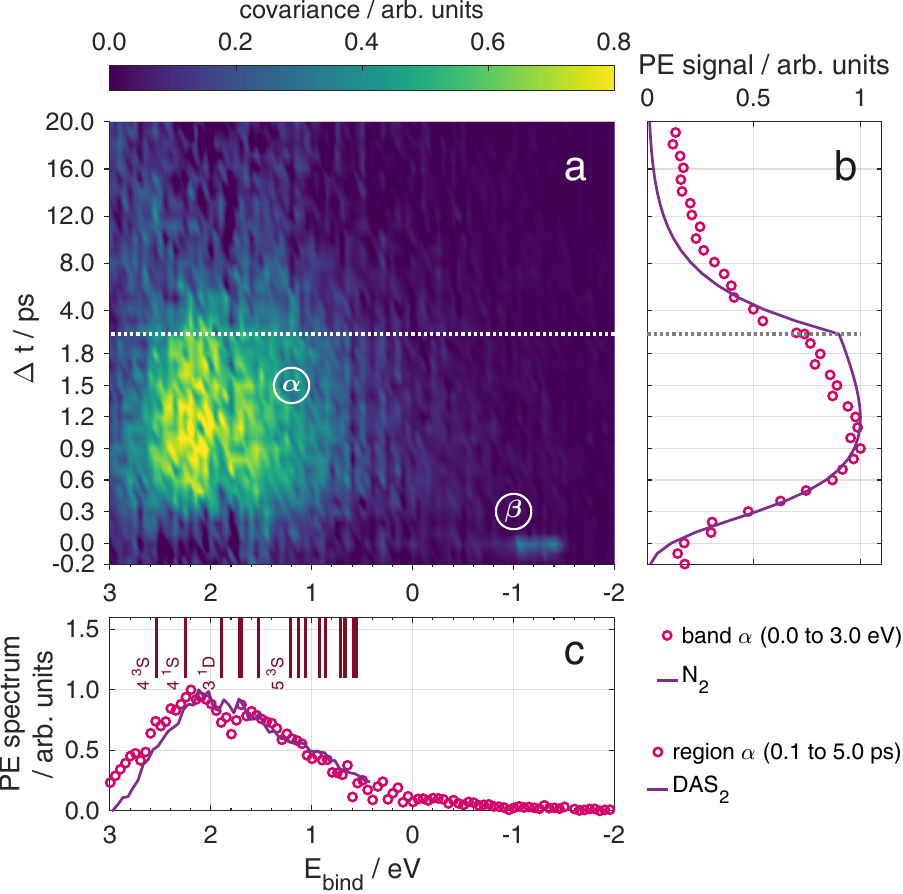}
\caption{Time-resolved photoelectron-photoion covariance spectrum of Mg$_n$ aggregates  inside \hen. 
\textbf{(a)} Pseudocolor plot of TRPES correlated to ion masses between 16 and 300\,u. Two energy bands \txtmath{(\alpha)} and \txtmath{(\beta)} are marked. 
\textbf{(b)} Comparison of energy-integrated covariance band \txtmath{(\alpha)} to the time-dependent decay function $N_2$ of \txtmath{DAS_2}.
\textbf{(c)} Comparison of the time-integrated covariance band \txtmath{(\alpha)} to \txtmath{DAS_2}. Red vertical lines indicate highly excited Mg atom states.\cite{NIST_ASD} 
\label{fig:timescan_covariance}
}
\end{figure}

\begin{figure}
    \centering
    \includegraphics[width=\linewidth]{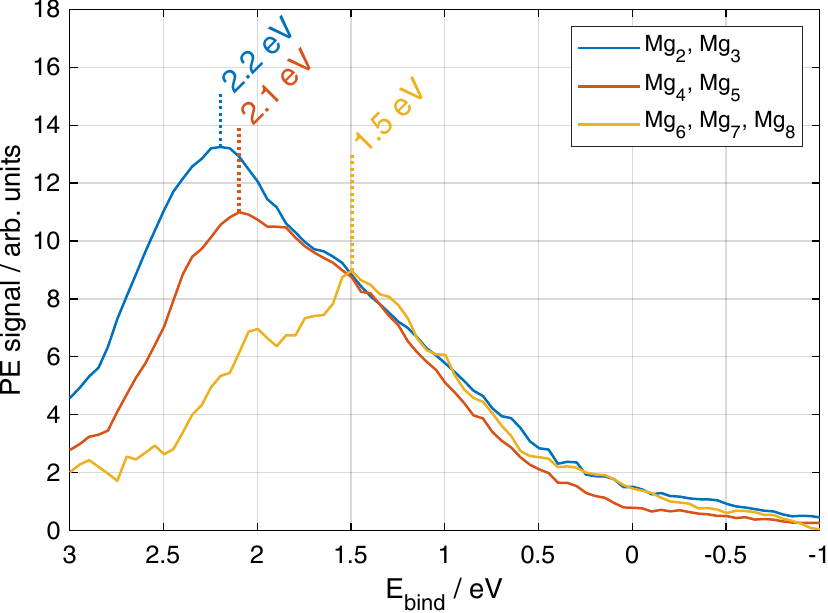}
    \caption{Time-integrated (0.1 ps to 5 ps) photoelectron spectra correlated with different \txtmath{Mg^+_nHe_m} ion complexes, as indicated in the legend. For the presentation, a moving mean with a width of 0.4~eV is applied. }
    \label{fig:PEspectra_different_cluster_sizes}
\end{figure}

\label{sec:Comparison DAS Cov.}
Figure~\ref{fig:timescan_covariance} shows the time-resolved PE spectra detected in covariance with $\mathrm{Mg_n^+}$ ($n=1-12$). 
In contrast to the transient PE spectrum of all electrons in Fig.~\ref{fig:timescan_all_electrons}, only two bands show up: A broad band \txtmath{(\alpha)} that extends from 0.5 to 3\,eV with a slow rise and fall time, and the cross-correlation band \txtmath{(\beta)} originating from three-photon ionization of gas phase Mg atoms [corresponding to band (4) in Fig. \ref{fig:timescan_all_electrons}].
We compare covariance band \txtmath{(\alpha )} to the global fit results of TRPES band (2), separately in the spectral domain (Fig.~\ref{fig:timescan_covariance}c) and in the temporal domain  (Fig.~\ref{fig:timescan_covariance}b).
This comparison reveals good agreement in both domains, which is a strong indicator that both observables reflect the same underlying photoinduced process, especially under consideration of the difference of the experimental methods. The fact that only band (2) of the TRPES is apparent in the covariance spectrum, while band (1) is missing, shows that only the photoexcitation process represented by band (2) leads to ejection of Mg$^+$ ions.
In contrast to the dilute aggregate,
photoexcitation of compact Mg$_\textrm{n}$ cluster (\txtmath{DAS_1}) leads to immediate promotion of electrons into the detection window for one-photon ionization of the probe pulse [band (1) in Fig.~\ref{fig:timescan_all_electrons}]. Fragmentation caused by energy conversion to nuclear motion appears much less pronounced for the compact cluster so that ion ejection from the droplet is prevented, eliminating band (1) from the covariance spectrum (Fig.~\ref{fig:timescan_covariance}).
This finding further supports our hypothetical assignment that the two bands correspond to different species with different photoexcitation dynamics: compact \txtmath{Mg_n} cluster associated with  $\mathrm{DAS_1}$ and the foam-like configuration associated with $\mathrm{DAS_2}$ and covariance band \txtmath{(\alpha ).} 
\\
An assignment of the electron signal shown in Fig.~\ref{fig:timescan_covariance} to specific cluster sizes provides additional insight into the fragmentation process.
To this end, Fig.~\ref{fig:PEspectra_different_cluster_sizes} shows time-integrated (0.1~-~5\,ps) electron spectra correlated to the  ion complexes \txtmath{Mg^+_{2,3}He_m}, \txtmath{Mg^+_{4,5}He_m} and \txtmath{Mg^+_{6,7,8} He_m}, with $m=0-5$. 
The most likely energy,  marked  by vertical lines in Fig. \ref{fig:PEspectra_different_cluster_sizes}, decreases from 2.2~eV for \txtmath{Mg^+_{2,3}He_m} to 1.5~eV for \txtmath{Mg^+_{6,7,8}He_m}. This development shows that  electronic relaxation to energetically lower states yields smaller fragments,  as more electronic energy is converted to nuclear kinetic energy.

\section{Discussion}
The combination of global fitting analysis of PE spectra and electron-ion covariance detection provides evidence that He droplets enable the formation of two Mg$_\textrm{n}$ configurations.  
These two configurations, represented by bands (1) and (2) in Fig.~\ref{fig:timescan_all_electrons}, show distinctively different photoexcitation responses, which is identified through the transient population of states with binding energies below 3.07\,eV, corresponding to the one-photon detection window of the probe-pulse.
We now discuss the energetics of the excitation processes leading to the population of these low-binding-energy states.
Concerning band (1), compact Mg$_\textrm{n}$ cluster have a lower ionization potential than the single Mg atoms and since calculated absorption spectra of small Mg$_\textrm{n}$ cluster~\cite{Solovyov2004, Shinde2017} overlap with our pump  photon energy, efficient pump-probe photoionization can be expected (Fig.~\ref{fig:schematic}c).
The energetics leading to band (2) are less obvious: The transition from foam-like aggregate to compact cluster is triggered by the 3$^1$P$_1$ atom excitation, which has a binding energy of 3.30\,eV and thus lies outside the one-photon detection window.
The highly excited state population of band (2) in the 0-3\,eV  binding energy range appears delayed, with a rise time of \foamrisetime~ (see Fig.~\ref{fig:timescan_all_electrons} and band ($\alpha$) in Fig.~\ref{fig:timescan_covariance}).
This raises the question about the processes leading to population of these low-binding-energy states up to the ionization continuum.
A further question that will be discussed in the following is the different fragmentation behavior of the two Mg$_\textrm{n}$ configurations in terms of ion ejection from the droplet. 
\\
\\
\textbf{Energy-pooling reaction.} 
The observed population of excited states above the pump photon energy  is reminiscent of energy pooling reactions, observed in mixtures of metal vapor and a noble gas~\cite{Kopystynska1982}. For Mg, 457.1\,nm excitation of two atoms from the 3$^1$S$_0$ ground state to the 3$^3$P$_1$ excited state at 4.94\,eV binding energy in a He buffer gas environment leads to population of the higher states 3$^1$P$_1$ and 4$^3$S$_1$ levels at 3.30\,eV and 2.54\,eV binding energy, respectively~\cite{Husain1986}. 
\\
In our experiment, aggregates of cold Mg atoms are photoactivated in a dilute configuration with a Mg--Mg distance of 9.5\,\AA~\cite{Hernando2008}. The excitation energy for the single--atom 3$^1$P$_1\leftarrow3^1$S$_0$ transition in this environment is slightly blue-shifted to 4.40\,eV photon energy (282.5\,nm), relative to the bare-atom transition at 4.35\,eV (285.5\,nm)~\cite{NIST_ASD}. The dynamics in the He solvation shell in response to the photoexcitation process~\cite{Thaler2018} leads to collapse of the dilute configuration and formation of a compact Mg$_\textrm n$ cluster. 
Interaction of ground-state and electronically excited Mg atoms leads to population of various states above the initially excited 3$^1$P$_1$ state, as shown by the photoelectron signal in Fig.~\ref{fig:timescan_covariance}c. 
While population of these states cannot result from a two-photon excitation process (see above), energy transfer through collisions of two excited 3$^1$P$_1$ atoms can populate all Mg states up to the ionization potential.
We thus propose a similar energy pooling reaction, based on the excited-state potential energy curves of Mg$_2$. \cite{Amaran2013,Knoeckel2014}. The reaction starts with the He-mediated formation of excited Mg$_2^*$ molecules~(see also Fig.~\ref{fig:schematic}c): 
\begin{align*}
    \mathrm{ Mg(3^1P_1) + Mg(3^1S_0) + He \rightarrow Mg_2^*(^1\Sigma^+_u, ^1\Pi_g) + He}
\end{align*}
Collision of Mg$_2^*$ with another excited Mg$^*$ atom leads to population of the highly-excited states: 
\begin{align*}
   \mathrm{ Mg_2^*(^1\Sigma^+_u, ^1\Pi_g) + Mg(3^1P_1) }\\\rightarrow 
   \mathrm{Mg(4^3S, 4^1S, 3^1D, ...) + 2 Mg(3^1S_0)}
\end{align*}
In the covariance spectrum shown in Fig.~\ref{fig:timescan_covariance}, the photoelectron band spanning from 0 to 3\,eV binding energy reveals the corresponding transient population distribution. On top of Fig.~\ref{fig:timescan_covariance}c, electronically excited states of Mg atoms are indicated for comparison. Since Mg$_\textrm{n}$ cluster  up to $n\sim20$ atoms exhibit non-metallic van der Waals-type bonding,~\cite{Diederich2005}  we refer to atomic states, which are, however, not resolved in the PE spectrum due to environmental broadening and laser bandwidth. 
\\
The proposed energy pooling reaction requires at least two excited Mg atoms in the foam-like aggregate. One can estimate the excitation probability of one Mg atom to be $p_1 = 0.81\pm 0.15$, based on the photon absorption cross section and photon density, including experimental uncertainties (see Appendix~\ref{subsec:excitation_probability}). Excitation of at least two Mg atoms is therefore quite likely. 
\\
\\
\textbf{Cluster formation and fragmentation dynamics.}
The characteristic rise time of the transient photoelectron signal representing the highly-excited state population [$N_2$ in Fig.~\ref{fig:timescan_all_electrons}b and band ($\alpha$) in Fig.~\ref{fig:timescan_covariance}b] represents the transition from a foam-like Mg aggregate  to a compact cluster. From parameters of the global fit population $N_2$, a time constant of \foamrisetime~ for cluster formation is determined. This value agrees with the characteristic time-constant of 350\,fs obtained by pump--probe strong-field ionization, which was proposed to represent the collapse of the dilute foam-like configuration \cite{Goede2013}.
\\
Concerning nuclear dynamics, including fragmentation of the Mg aggregate and acceleration of the fragments, it is important to realize that electronic energy can be converted into kinetic energy in each  step of the energy pooling reaction. Initially, He atoms carry away some of the 4.4\,eV photon energy stored in Mg$^*$ in order to form the Mg$_2^*$  bond. 
In step two, the Mg$_2^*$--Mg$^*$ collision can lead to the population of various higher excited Mg$^*$ states (4$^3$S, 4$^1$S, 3$^1$D, ...). The difference in excitation energies of reactants ($8.7$\,eV) and products is thus converted into Mg kinetic energy, ranging from $\leq$3.6\,eV for population of the 4$^3$S state to $\leq$1.9\,eV for population of states close to the IP (the ''$\leq$'' accounts for the kinetic energy of the He atoms).~\cite{NIST_ASD} 
The released kinetic energy will increase the cluster temperature and, since the amount of  energy is comparable to the binding energy of small Mg$_\textrm{n}$ cluster \cite{Jellinek2002}, fragmentation and ejection from the droplets are expected. Figure~\ref{fig:PEspectra_different_cluster_sizes} supports this assumption by showing that the population of lower electronic states is correlated with smaller clusters. This is in line with previous observations, where liberation of ions from the droplet (overcoming the solvation energy) is only possible through a release of  kinetic energy~\cite{Stadlhofer2022}. Note, however, that the ionization potentials decreases with cluster size~\cite{Kazak2022}. 
\\
The covariance measurements also reveal that the ion--to--electron ratio decreases from 52\,\% for gas-phase Mg atoms (a characteristic value of our covariance spectrometer)
to $\sim15$\,\% for Mg$_\textrm{n}$--He$_\textrm{N}$. This shows that the energy release is only in one out of three cases sufficient for liberation of the fragment ion from the solvation energy of the He droplet.
\\
\\
\textbf{Relaxation of excited cluster through electron--phonon interaction.}
The decaying character of the TRPES signal in Fig.~\ref{fig:timescan_all_electrons}, observed for both compact [band (1)] and foam-like aggregates [band (2) in Fig.~\ref{fig:timescan_all_electrons} and ($\alpha$) in Fig.~\ref{fig:timescan_covariance}], indicate on fast electronic relaxation.  
The characteristic decay time constant of the compact Mg$_\textrm{n}$ cluster is \compactdecaytime. 
Electronically excited metal cluster typically relax via electron--electron interaction on a time scale of less than $\sim100$\,fs.~\cite{Schirato2023,Stolow2004,Hertel1996}  
Our observation of a slower decay thus supports the assumption of a non-metallic complex~\cite{Young2009}, in line with recent studies showing small magnesium clusters ($n\le18$) are not metallic~\cite{Diederich2001, Thomas2002}.
The electronic relaxation in van der Waals clusters proceeds through non-adiabatic transitions to lower electronic states with vibrational excitation.  Electron--phonon coupling  thus leads to electronically relaxed but vibrationally hot clusters. 
\\
For the Mg$_\textrm{n}$ cluster formed from the foam-like configuration, the photoelectron signal decay reveals a ten times longer time constant of 
\foamdecaytime, compared to the compact cluster. This further points at the non-conductive van der Waals nature of Mg$_\textrm{n}$ clusters. The significantly slower decay might be rooted in the excitation of higher electronic states due to energy pooling, or a reduced electron--to--phonon energy transfer because the formed cluster are hot and potentially reduced in size due to fragmentation. Also, cluster (fragments) ejected from the droplet loose contact to the thermal bath, keeping them vibrationally excited for longer times.
The transformation of electronic to vibrational energy within $\sim10$\,ps leads to a reduction of the cluster fragment size, as depicted in the time-resolved mass spectra in Fig.~\ref{fig:timescan_all_ions_waterfall}. 
Electronic relaxation of the formed cluster also manifests as increase of the 3$\mathrm{^1P_1}$ population with a characteristic time of $\tau_{BG}^{low} = 3.3~\mathrm{ps}$ (see band (3) in Fig.~\ref{fig:timescan_all_electrons} and Appendix~\ref{subsec:Supplement:Global Fitting Method}).
The increasing \MgExcited population 
establishes agreement with the steady-state photoelectron spectra reported for dilute Mg$_\textrm{n}$ ensembles in \hen~\cite{Kazak2019}.
In photoemission experiments using nanosecond laser pulses, a strong signal from  \MgExcited and a relatively weak signal in the region above at 1--2\,eV binding energyis observed. Considering that excitation and ionization occurs with two photons within a 10\,ns time window, in combination with the picosecond lifetime of higher excited states and the increasing \MgExcited population, our time-domain observations are in agreement with the steady state results. \\
\section{Conclusions}
Femtosecond time-resolved photoelectron and -ion spectroscopy and a subsequent global fitting analysis has been used to study the light-induced dynamics of small magnesium clusters with emphasis on foam-like complexes. Photoexcitation of this metastable Mg$_\textrm{n}$  configuration leads to the contraction of the aggregate on a picosecond timescale. 
The contraction initiates the transient population of highly excited Mg states through energy pooling, as well as pronounced nuclear dynamics, which  can clearly be resolved and distinguished from the compact Mg cluster response by inspecting the transient photoelectron signals.
\\
The crucial prerequisite  for this observation is the stabilization of Mg atoms at nanometer interatomic distance inside superfluid helium. 
First attempts to simulate these exceptional solvation properties of Mg atoms in \hen ~have led to contradictory results: While static DFT simulations predicted stable separation of two Mg atoms at 9.5\,\AA~distance in He~\cite{Hernando2008}, path integral Monte Carlo simulations found  only equilibration to strongly bound Mg$_2$ and Mg$_3$ molecules~\cite{Krotscheck2016}. Clarification of this discrepancy with frequency-domain spectroscopy might be challenging because the absorption spectra of small Mg$_\textrm{n}$ cluster are predicted to show a small dependence on the cluster size~\cite{Solovyov2004,Shinde2017}, with many of them  absorbing at the 4.4\,eV transition of the metastable configuration~\cite{Przystawik2008}.
Our time-domain analysis provides additional insight by revealing the photodynamical response: 
Photoexcitation of the foam-like Mg$_\textrm{n}$ configuration triggers nuclear dynamics leading to the contraction of the aggregate. This nanometer motion of Mg atoms is represented by significantly slower transient photoelectron signals, compared to the predominantly  electronic dynamics of a compact van der Waals cluster. 
This difference in the transient response of the two Mg$_\textrm{n}$ configurations is the essential ingredient for distinguishing the overlapping spectra through global fitting analysis. Purely based on frequency-domain information, this distinctiveness is not given.
\\
The observation that both configurations are  simultaneously present within the observed Mg$_\textrm{n}$He$_\textrm{N}$ ensemble is of relevance for cluster formation inside He droplets \cite{Ernst2021}.
While the foam-like configuration is predicted to be favorable under steady-state conditions at the droplet temperature~\cite{Hernando2008}, this weakly-bound aggregate can collapse during the pickup phase either by  a hot (insufficiently cooled) Mg atom or when the dilute aggregate exceeds a critical size~\cite{Kazak2022}.
Cluster aggregation inside \hen, in particular the formation of foam-like configurations, is thus governed by an interplay of kinetics and  thermalization. 
Recent time-dependent simulations, following TD-DFT~\cite{Coppens2019a, GarciaAlfonso2022, Fixot2024, Trejo2024}, particle-based~\cite{GarciaAlfonso2022, Fixot2024}, or hybrid~\cite{BlancafortJorquera2019} approaches  are able to account for such kinematic effects. These simulations reveal that the growth of compact clusters can be hindered by freezing in metastable configurations, with a certain probability depending on kinematic parameters.
\\
Metastable separation through formation of a He barrier has been predicted, in addition to Mg~\cite{Hernando2008}, for rare-gas~\cite{Eloranta2008,GarciaAlfonso2022,Coppens2019a,BlancafortJorquera2019,Fixot2024,Trejo2024} and halogen~\cite{Eloranta2011} atoms, as well as larger molecules~\cite{Calvo2016}.
Experimental evidence for the existence of a metastable configuration was reported in early deflection and mass spectrometric experiments of Ar, Kr, Xe, H$_2$O and SF$_6$, which  find that the cross section for pick up is larger than that for coagulation~\cite{Lewerenz1995}.
Electronic spectroscopy of anthracene--Ar cluster in \hen ~revealed indications for the shielding of an attached Ar atom by a helium layer
\cite{Lottner2020,Calvo2021}.
Very recently, electron diffraction also found evidence for large Xe-Xe distances with He located in between~\cite{Trejo2024}.
In experiments with bulk liquid He, a very similar stabilization of atoms was reported~\cite{Gordon1974}. Cold impurity atoms introduced into the He solvent through a supersonic jet expansion are found to condensate in an ''impurity--helium solid'', characterized by a pronounced spatial separation of the impurities~\cite{Gordon2004}. Such structures, investigated by means of optical spectroscopy, electron spin resonance and thermometry, were recently also observed for H$_2$O clusters~\cite{Efimov2013, MezhovDeglin1999}. 
\\
While the ability of \hen~to freeze aggregates in non-equilibrium structures  has long been appreciated~\cite{Nauta1999a}, stable and well-defined large-distance separation of reactants through a solvent layer barrier provides fundamentally new perspectives for bond-formation studies. 
Helium droplets furthermore enable stable separation of surface-located and solvated species~\cite{Lackner2018,Poms2012} and the possibility to switch between the two locations through electronic excitation~\cite{Kautsch2015} or through ionization, which recently enabled the real-time observation of the primary steps of ion solvation in helium~\cite{Albrechtsen2023}.
Taking into account the formation of exciplexes, consisting of excited atoms and He~\cite{Reho1997,Schulz2001,Droppelmann2004, Ziemkiewicz2015}, shows that helium nanodroplet isolation holds great promise to study  bond formation dynamics in various species.
   Such studies will provide insight into elementary processes accompanying the photoinduduced formation of chemical bonds, such as the transient population of highly excited states above the excitation photon energy, as observed here for the formation of Mg clusters. 
The proposed energy pooling process relies on merging the energy of two or more excited Mg atoms to populate highly excited states. The ability of photon upconversion to trigger photoinduced processes that lie outside the available spectrum has implications in various fields, including photomedicine~\cite{Bolze2017, Haase2011}.
Efficient upconversion requires close distances of the involved particles, as recently demonstrated with solid-state organic chromophore blends~\cite{Weingarten2017}, whereas  gas phase configurations suffer from prohibitory low yields due to large particle distance.\cite{Kopystynska1982,Husain1986}
The nanometer confinement provided by He droplets, together with flexible opportunities for generating tailor-made aggregates, provides thus a new and promising route to characterize the underlying energy and charge-transfer dynamics. 
\section{Acknowledgments}
We thank Wolfgang E. Ernst for useful discussions. This research was funded in whole or in part by the Austrian Science Fund (FWF) [10.55776/P33166]. For open access purposes, the authors have applied a CC BY public copyright license to any author-accepted manuscript version arising from this submission.
The authors acknowledge support from NAWI Graz.
MS acknowledges funding
as recipient of a DOC Fellowship (26387) of the Austrian Academy of Sciences at the Institute of Experimental Physics.
JT acknowledges the Deutsche Forschungsgemeinschaft (TI 210/13-1  and SFB\,1477 'Light-Matter Interactions at Interfaces,' Project No. 441234705) for financial support. 
\\

%

\newpage
\section{Appendix}
\subsection{Generation and doping of He droplets}
\label{subsec:doping}
\begin{figure}
    \centering
    \includegraphics[width= \linewidth]{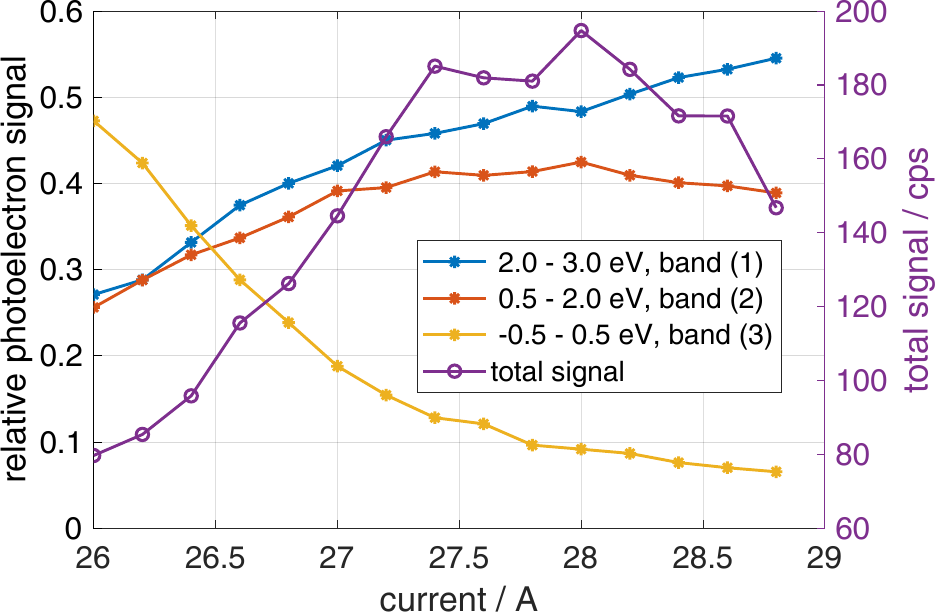}
    \caption{ Dependence of the pump-probe photoelectron bands (c.f., Fig. ~\ref{fig:timescan_all_electrons}) on the Mg doping level, represented as heating current applied to the Mg pickup oven. The photoelectron spectra for different heating currents are normalized and the depicted photoelectron signal is obtained through integration within the energy intervals given in the legend, and in time from -1~ps to 11~ps.}
    \label{fig:pump_probe_pickup_scan}
\end{figure}

We generate He droplets by expanding high-purity He through a 5~\txtmath{\mu m} diameter nozzle at 20~bar pressure of  and  11.6 K temperature. The droplet size follows a log-normal distribution with a  mean number size of $\overline{N} = 13500$  \cite{Toennies2004}. The He droplets are doped with Mg atoms inside a resistively heated oven, where the number of Mg atoms entering the droplet can be controlled by changing the metal vapor pressure via the oven heating current. 
A quadrupole mass spectrometer is used to monitor the Mg pickup conditions.
The \txtmath{Mg_n} aggregate size limit can be estimated from the mass-to-charge spectrum in Fig. \ref{fig:timescan_all_ions_waterfall}, where \txtmath{Mg_8^+} corresponds to the highest detected mass-to-charge ratio.
By assuming that the parent cluster was split in half during pump--probe ionization, the largest initial cluster size can be estimated with \txtmath{Mg_{16}}.
To characterize the influence of the Mg doping level on bands (1)--(3)  in the photoelectron spectrum (Fig. \ref{fig:timescan_all_electrons}), the corresponding photoelectron signals are shown in  Fig. \ref{fig:pump_probe_pickup_scan} as a function of the Mg oven heating current.
While band (1) (2 to 3~eV binding energy) increases steadily with rising oven current, band (2) (0.5 to 2~eV binding energy) shows a local maximum at around 28~A. This opposing signal dependency for large doping levels supports the assignment of band (1) to compact \txtmath{Mg_n} clusters and band (2) to a foam-like \txtmath{Mg_n} configuration and is in line with the recent observation of spontaneous foam collapse for higher Mg doping levels \cite{Kazak2022}.  PE band  (3), associated with single Mg atoms (-0.5 to 0.5~eV binding energy)  is most prevalent at lower oven currents, and is strongly suppressed for larger oven currents, confirming that the He droplets contain multiple Mg atoms for the applied heating current of 29\,A.
\subsection{Global fitting}
\label{subsec:Supplement:Global Fitting Method}
\begin{figure}[ht]
    \centering
    \includegraphics[width=\linewidth]{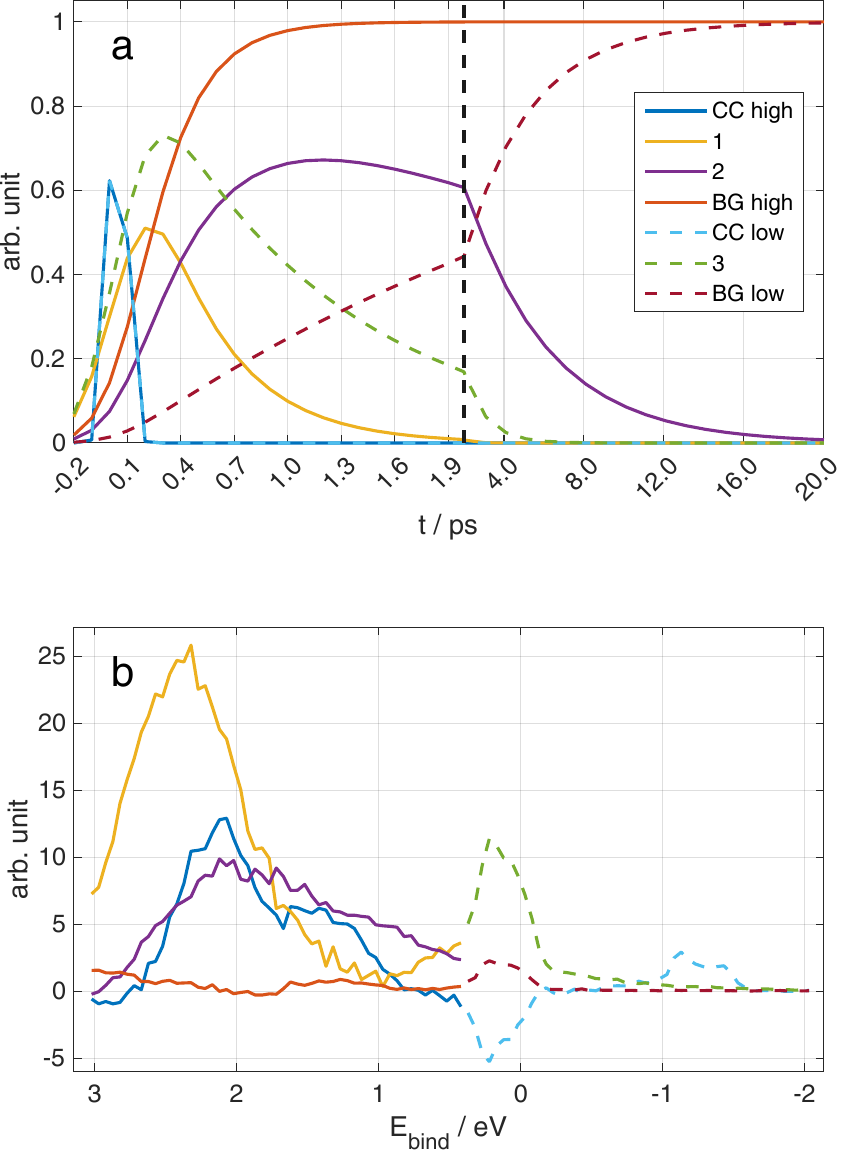}
    \caption{Overview of the global fit analysis.
    \textbf{(a)} Time-dependent decay functions for each decay associated spectrum. Note that the time-scale changes at 2~ps, indicated with the black-dashed vertical line. The crosscorrelation functions are abbreviated as CC, and the background functions as BG. Background and crosscorrelation also have the identifier high and low, which refers to  the low or high binding energy region (see main text). 
    \textbf{(b) }Decay associated spectra corresponding to the decay functions in (a). }
    \label{fig:L2_populations_and_spectra}
\end{figure}

\begin{figure}[ht]
    \centering
    \includegraphics[width=\linewidth]{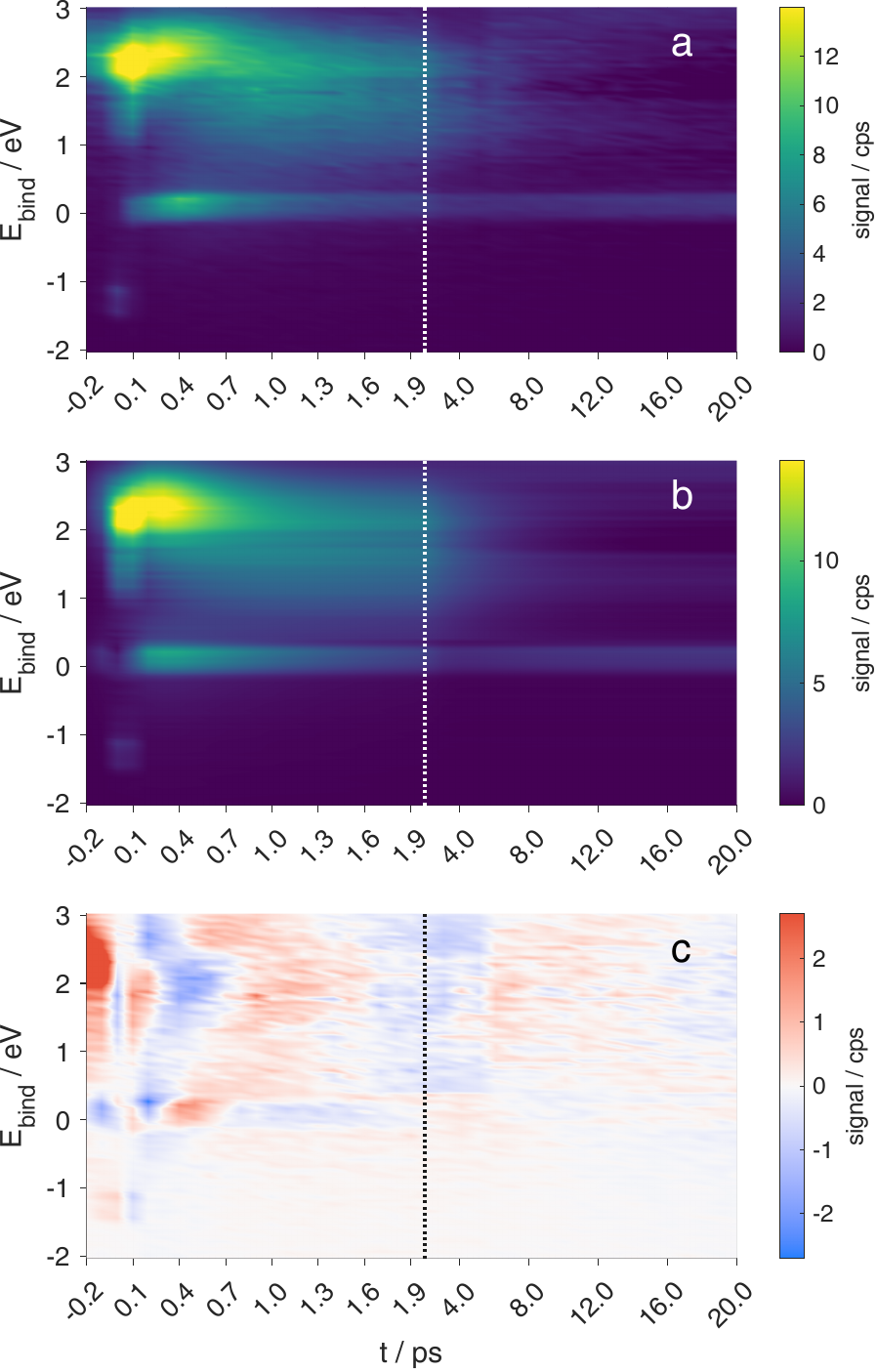}
    \caption{ Evaluation of the global fit model. 
   \textbf{ (a)} Measured  time-resolved photoelectron spectrum as depicted in Fig.~\ref{fig:timescan_all_electrons}a 
    \textbf{(b)} Reconstruction of the time-resolved photoelectron spectrum, obtained by summing up all DAS contributions. 
    \textbf{(c)} Residual plot obtained as difference of the measured (a) and reconstructed (b) spectra.}
    \label{fig:L2_surf}
\end{figure}

For our global fitting analysis, the following two types of transient decay functions are used:
(i) An instantaneously rising  signal followed by exponential decay is modelled by a decay function of the form
\begin{equation}
\label{eq:DAS_S1}
    N_i(t) = \frac{1}{2} e^{\frac{\sigma ^2}{2\tau_i^2} -  \frac{(t - t_0)}{ \tau _i}} \text{erfc}\left(-\frac{t - t_0 }{\sqrt{2} \sigma }+\frac{\sigma }{\sqrt{2} \tau _i}\right),
\end{equation}
which is obtained through convolution of a Gaussian function  with a full halfwidth $\sigma$ (representing the temporal instrument response function) centered at time zero, $t_0$, and a single exponential decay function with characteristic decay time $\tau_i$~\cite{Stokkum2004}.
In Eq.~\ref{eq:DAS_S1} erfc is the complementary error function 
\textrm{erfc}(z) = 1 - \textrm{erf}(z).
\\
(ii) Modeling of a delayed signal rise followed by an exponential decay is achieved through a sequential model, in which state A is excited by the pump pulse  and decays with time constant $\tau_A$ into state B,  which itself subsequentially decays with time constant $\tau_B$. The corresponding decay function reads
\begin{equation}
\label{eq:DAS_S2}
    N_i(t) = [N_A(t ,\tau_A) - N_B(t,\tau_B)] \frac{\tau_A}{2 ( \tau_A - \tau_B )},
\end{equation}
where $N_A(t ,\tau_A)$ and $N_B(t ,\tau_B)$ are represented by Eq.~\ref{eq:DAS_S1}. The time constant $\tau_A$ thus corresponds to the characteristic rise time of the population $N_i(t)$, labeled $\tau_i^\textrm{rise}$ in the main text, whereas $\tau_B$ corresponds to the characteristic decay time of $N_i(t)$, labeled $\tau_i$.
In addition to the three DAS and corresponding decay functions discussed in the main text, in both energy regions an exponentially increasing background is considered by using Eq.~\ref{eq:DAS_S2} with a corresponding rise time and a very long decay time of 1~$\mathrm{\mu s}$.
The fitting procedure reveals that the  background in the low binding energy region has a rise time of $\tau_{BG}^{low} =(3300\pm1200)$~fs 
and in the higher binding energy region has a rise time of $\tau_{BG}^{high} = (200 \pm 1200)$~fs. 
\\
There is still some interplay between \txtmath{DAS_3} and crosscorrelation in the low binding energy region, which suggests that $N_3$ might rise slightly slower than the single decay fit function allows with the temporal instrument response of $\sigma = 170$~fs. However, switching to a two-level decay fit function does not improve the fit for band (3).
Figure~\ref{fig:L2_populations_and_spectra}a summarizes all decay functions, and the corresponding decay associated spectra are shown in Fig.~\ref{fig:L2_populations_and_spectra}b.
In Fig.~\ref{fig:L2_surf} the quality of the fit result can be evaluated by comparing the measured time-resolved photoelectron spectrum in (a) to the global fit model in (b). The residuals in (c) show that the deviation between the two is below 10 \% throughout the majority of the spectrum.
\subsection{Excitation probability for Mg atoms inside helium droplets}
\label{subsec:excitation_probability}
The energy pooling process outlined in the main text requires at least two excited Mg atoms inside a droplet. We therefore estimate the excitation probability of the Mg atoms for the applied laser parameters. The probability $  p_\textrm{abs}$ of a single photon from the pump laser pulse to be absorbed by one Mg atom is given by the ratio of absorption cross section $\sigma_\textrm{abs}$ and the laser beam area $A$: 
\begin{equation}
\label{eq:absorption_probability}
     p_\textrm{abs}= \frac{\sigma_\textrm{abs}}{ A} 
\end{equation}
\vspace{7mm}
\textbf{Calculation of the absorption cross section $\sigma_\textrm{abs}$}\\
Since the width of the absorption spectrum of Mg atoms solvated inside He droplets \cite{Przystawik2008} is comparable to the spectral width  of the short laser pulses, the absorption cross section $\sigma_\textrm{abs}$ can be calculated by multiplying the frequency-integrated absorption cross section $\sigma_{0}$ with the overlap integral $I_\textrm{overlap}$ of the line shape function of the broadened atom transition $g(\omega)$  and the spectral line shape of the laser pulse $\rho(\omega)$:
\begin{equation}
    \sigma_\textrm{abs}= \sigma_0  \times I_\textrm{overlap}
    = \sigma_0\int_{-\infty}^{\infty}  g(\omega) \rho(\omega)  \mathrm{d}\omega 
\label{eq:absorption_prob}
\end{equation}

Note that $g(\omega)$ and $\rho(\omega)$ are in units of $\mathrm{s/rad}$.
The frequency-integrated absorption cross section $\sigma_{0}$ for the Mg atom transition $3^1P_1 \leftarrow 3^1S_0$  follows from  the Einstein $A_{21}$ coefficient~\cite{Hilborn1982}:
\begin{equation*}
    \sigma_{0}= \frac{1}{4}\frac{g_2}{g_1} \lambda^2 A_{21}  
\end{equation*}
Here $g_2= 3$ and $g_1= 1$ are the degeneracies in the upper and lower level, respectively, and $\lambda$ is the excitation wavelength.
With the Einstein spontaneous emission rate for Mg of~\cite{Kelleher2008}
\begin{equation*}
    A_{21} = 4.91\times 10^8 \mathrm{s^{-1}}
\end{equation*}
and the  $3^1P_1 \leftarrow 3^1S_0$ transition wavelength of Mg atoms in He droplets of~\cite{Przystawik2008}
\begin{align*}
     \lambda = 282.5 \, \mathrm{nm,} 
\end{align*}
we obtain the frequency-integrated absorption cross section $\sigma_{0}$ of 
\begin{equation*}
    \sigma_{0} = 2.90389\times10^{-5} \mathrm{~m^2 }\frac{\mathrm{rad} }{\mathrm{s } }
\end{equation*}
In using this value, we are assuming that the spontaneous emission rate for Mg does not change when the atoms are inside the helium droplet i.e. that the broadened lineshape  is not caused by a shorter lifetime of the excited state. \\
For the spectral line shape of the laser pulse $\rho(\omega)$, a Gaussian spectrum with standard deviation $s_\textrm{laser}$ centered at the mean angular frequency $\omega_0$ is assumed: 
\begin{equation*}
\rho (\omega) = \frac{1}{\sqrt{2 \pi} s_\textrm{laser}} e^{ -(\omega - \omega_{0})^2/(2s_\textrm{laser}^2)}   
\end{equation*}
From the measured spectrum of the pump pulses, with a full width at half maximum (FWHM) bandwidth of $\Delta \rho = 3.5~$nm centered at  $282.5~$nm,  the standard deviation of the laser spectrum in units of angular frequency can be calculated:
\begin{equation*}
s_{\mathrm{laser}}  =  3.51\times 10^{13} \,\mathrm{rad/s}
\end{equation*}

For the absorption spectrum of Mg in helium droplets,  we also assume a Gaussian line shape with a FWHM of $\Delta g = 4~$nm,~\cite{Przystawik2008} yielding a standard deviation of the line shape function of

\begin{equation*}
    s_{ \mathrm{abs}} = 4.01\times 10^{13} \, \mathrm{rad/s.}
\end{equation*}
With the spectral distributions of the laser and absorption line, we can calculate the  overlap integral:
\begin{equation*}
    I_\textrm{overlap}  = 7.488\times10^{-15} \, \mathrm{s/rad}
\end{equation*}

The final cross section according to Eq. \eqref{eq:absorption_prob} is then
\begin{equation*}
    \sigma_\textrm{abs} =   2200 \mathrm{~Mb}.
\end{equation*}

\vspace{7mm}
\textbf{Calculation of the laser beam area.}\\
For a Gaussian beam, the area $A(z)$ and diameter $d(z)$ as function of the distance $z$ to the beam waist are given by the following relations~\cite{Saleh2007}:
\begin{equation*}
    A(z) = \pi \left(\frac{d(z)}{2}\right)^2
\end{equation*}
\begin{equation*}
    d(z) = d_0 \sqrt{1 + \left(\frac{z}{z_R}\right)^2}
\end{equation*}
\begin{equation*}
    z_R = \frac{\pi d_0^2}{4 \lambda}
\end{equation*}
\begin{equation*}
    d_0 = 
    \frac{4  \lambda f M^2}{\pi D},
\end{equation*}
where $f$ is the focal length of the lens, $\lambda$ is the laser wavelength, $M^2$ is the beam quality and $D$ is the diameter of the laser beam at the lens. 
With the laser beam parameters listed in Tab.~II,  we obtain the following value for the laser focus area:
\begin{equation*}
    A = 4.63 \times 10^{14}~\mathrm{Mb}
\end{equation*}

\begin{table}
\label{tab:laser}
\caption{Laser beam parameters: $E_\textrm{pulse}\ldots$ pulse energy, $D\ldots$  diameter of the laser beam at the lens, $z\ldots$\ distance to the beam waist, $f\ldots$ focal length of the lens, $M^2\ldots$ beam   quality factor. }
\begin{tabular}{l|l|l|l|l|l}
$E_\textrm{pulse}$ / $\mu$J & $D$ / mm & $z$ / mm & $f$ / mm& $M^2$  \\\hline
1 & 2 & 0 & 1000 & 1.35  \\
\end{tabular}
\end{table}
\vspace{5mm}
\textbf{Calculation of the excitation probability.}\\
The single photon absorption probability of a Mg atom inside a He droplet, calculated according to Eq. \eqref{eq:absorption_probability}, is:
\begin{equation*}
    p_\textrm{abs} = \frac{2.2 \times 10^{3} ~\mathrm{Mb} }{ 4.63 \times 10^{14} ~\mathrm{Mb}} = 4.75\times10^{-12}
\end{equation*}

\begin{figure}[ht]
    \centering
    \includegraphics[width = \linewidth]{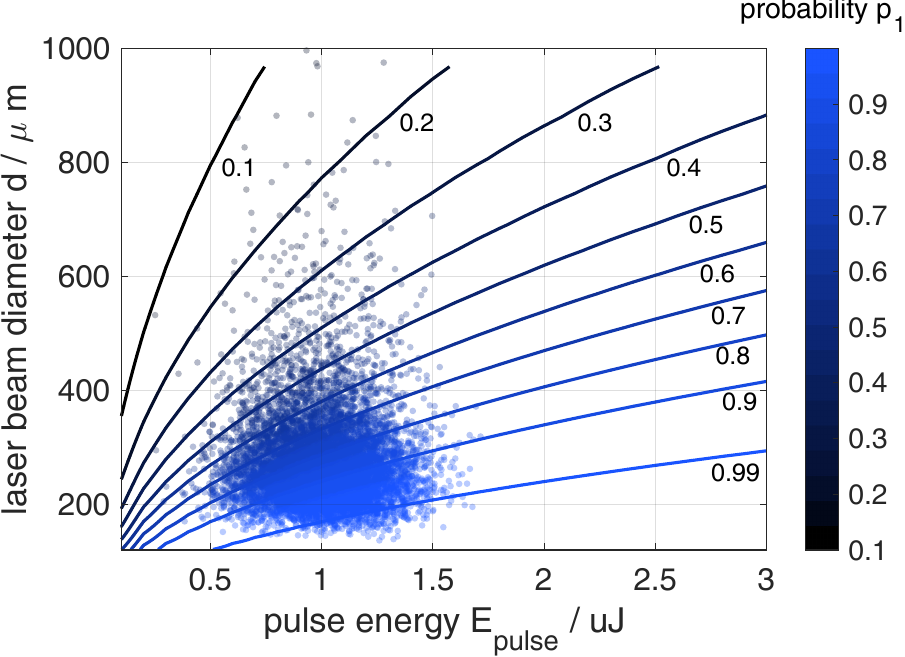}
    \caption{Sampled probabilities $p_1$, for one Mg atom to be excited by $N$ photons. The contours and colors are used to indicate the excitation probability for a given laser beam diameter $d$  and pulse energy $E_\textrm{pulse}$.  }
    \label{fig:p_1_random_samples}
\end{figure}

For the $N$ photons contained in a laser pulse, the excitation process can be modeled by a Bernoulli trial, i.e., the probability that at least one photon excites the atom follows a Bernoulli distribution:
\begin{equation}
\label{eq:bernulli}
    p_{1} = \sum_{n=0}^{N-1} p_\textrm{abs}\times (1-p_\textrm{abs})^n = 1 - (1-p_\textrm{abs})^N
\end{equation}
The number of photons $N$ in the pulse can be calculated from the pulse energy $E_\textrm{pulse}$ and photon energy ($\hbar \omega_0$ neglecting the bandwidth of the laser):
\begin{equation*} 
    N = \frac{E_\textrm{pulse}}{\hbar \omega_0}
\end{equation*}
For the value of $p_\textrm{abs} = 4.75\times 10^{-12}$, pulse energy of $1~\mathrm{\mu J}$ and photon energy of $h\times 1.061\cdot10^{15}$~Hz,  Eq. \eqref{eq:bernulli} yields a pump-pulse  excitation probability of a single Mg atom inside a He droplet of
\begin{equation*}
    p_1 = 0.9988.
\end{equation*}
Since this value is derived for optimal laser beam parameters, which might deviate in the experiment, we assume a Gaussian distribution of the parameters $E_\textrm{pulse}$, $D$ and $z$ (see Tab.~\ref{tab:probability_random_sampling}) and apply statistical sampling to generate a distribution for the excitation probability, from which we determine the mean and standard deviation. Large uncertainties of the parameters are chosen in order to obtain  a robust estimate for the excitation probability.
For each set of random sample parameters, an excitation probability  $p_1$ is calculated. In Fig.~\ref{fig:p_1_random_samples}, the pulse energy $E_\textrm{pulse}$ and laser beam diameter $d$ (at the sample) are chosen to depict the dependency of the excitation probability  in a 2D plot, together with the contour lines of the $p_1(d, E_\textrm{pulse})$ function.

\begin{table}

\caption{Mean and standard deviation for the parameters used in random sampling:  $E_\textrm{pulse}\ldots$ pulse energy, $D\ldots$ laser beam diameter at the focusing lens, $z\ldots$ sample position with respect to the laser beam waist. \label{tab:probability_random_sampling}}
\begin{tabular}{l | l | l | l}
& $E_\textrm{pulse}$ / $\mathrm{\mu J}$& $D$/ mm & $z$ / mm \\\hline
mean & 1.0 & 2.0 & 0 \\\hline
standard deviation & 0.2 & 0.5 & 50 \\
\end{tabular}
\end{table}

\begin{figure}
    \centering
    \includegraphics[width = \linewidth]{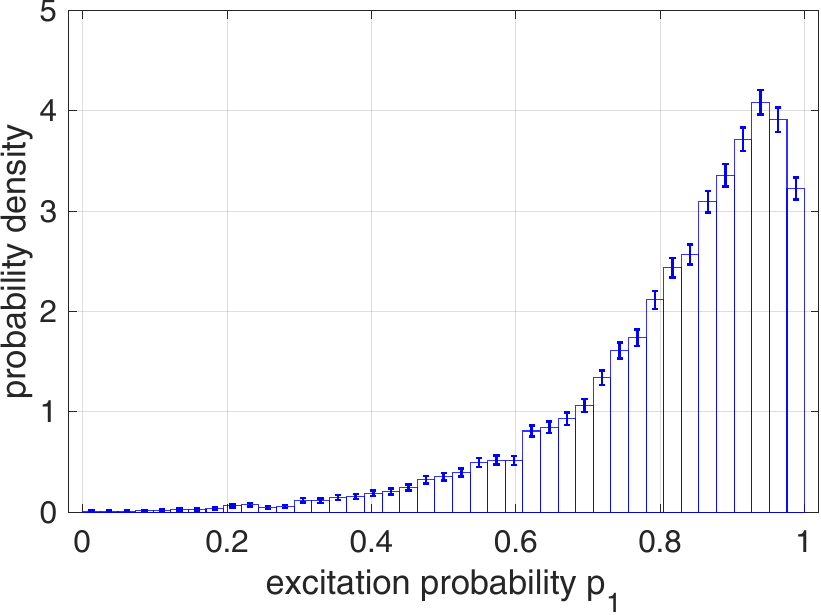}
    \caption{Numerical probability density for the excitation probability $p_1$, given the uncertainty in pulse energy, beam diameter at the lens and distance from sample in Table \ref{tab:probability_random_sampling}. }
    \label{fig:p_1_distribution}
\end{figure}

Fig. \ref{fig:p_1_distribution} shows a histogram of the sampled probabilities $p_1$, which can be interpreted as a  probability density for the excitation probability $p_1$, resulting from the distributions of input parameters.
The numerical mean and standard deviation of this distribution are 
    $p_1 = 0.81$ and $\sigma_{p_1} = 0.15$,
showing that Mg atom excitation inside He droplets is very likely for a broad range of experimental laser parameters. The high single-atom excitation probability makes excitation of multiple atoms inside a single droplet plausible. 

\end{document}